\documentclass[letter]{elsarticle}

\usepackage{lineno,hyperref}
\modulolinenumbers[5]

\usepackage{graphicx}        % standard LaTeX graphics tool
\newcommand{\be}{\begin{equation}}
\newcommand{\ee}{\end{equation}}
\newcommand{\bra}{\langle}
\newcommand{\ket}{\rangle}
\newcommand{\bea}{\begin{eqnarray}}
\newcommand{\eea}{\end{eqnarray}}

\journal{Journal of \LaTeX\ Templates}

\biboptions{authoryear}

\begin{document}

\begin{frontmatter}

\title{Rough volatility of Bitcoin}

%% Group authors per affiliation:
\author{Tetsuya Takaishi\corref{mycorrespondingauthor}}
\cortext[mycorrespondingauthor]{Corresponding author}
\address{Hiroshima University of Economics, Hiroshima 731-0192 JAPAN}
%%\cortext[mycorrespondingauthor]{Corresponding author}
\ead{tt-taka@hue.ac.jp}

\begin{abstract}
Recent studies have found that the log-volatility of asset returns exhibit
roughness.
This study investigates roughness or the anti-persistence of Bitcoin volatility. 
Using the multifractal detrended fluctuation analysis, 
we obtain the generalized Hurst exponent of the log-volatility increments
and find that the generalized Hurst exponent is less than $1/2$,
which indicates log-volatility increments that are rough.
Furthermore, we find that the generalized Hurst exponent is not constant.
This observation indicates that the log-volatility has multifractal property.
Using shuffled time series of the log-volatility increments, we infer that 
the source of multifractality  partly comes from the distributional property.
\end{abstract}


\end{frontmatter}

%{\bf Highlights}
%\begin{itemize}
%\item
%We study the log-volatility increments of Bitcoin.
%\item
%We not only find roughness but also multifractality in the log-volatility increments.
%\item
%The temporal correlations in the time series contribute to roughness. 
%\item
%The source of multifractality in the log-volatility increments partly comes from the distributional property.
%\end{itemize}

%\linenumbers

\section{Introduction}

Studies have intensively examined the statistical properties of asset prices and confirmed the existence of universal properties across various asset returns.
These properties are now classified as ``stylized facts,'' which include
(i)fat-tailed distributions, (ii)volatility clustering
(iii) slow decay of autocorrelation in absolute returns, 
and so on, see for example, \citet{Cont2001QF}.
The stylized fact (iii) also characterizes volatility to be long memory, and
more generally, the power transformed absolute returns $|r_t|^d$ have high autocorrelation for long lags\citep{taylor1986modelling}.
The power $d$, which gives the highest autocorrelation, is dependent on assets, 
and the autocorrelation is highest for stocks when $d$ is around 1\citep{ding1993long}.
For other assets, see, 
\citet{granger1995some,ding1996modeling,genccay2001introduction,takaishi2018taylor}.

It is important to model volatility with these properties to estimate 
or forecast an accurate volatility value, such as option pricing, and risk management of assets.
The most successful volatility models might be the autoregressive conditional heteroscedasticity(ARCH)\citep{Engle1982autoregressive}
and generalized ARCH (GARCH)\citep{Bollerslev1986JOE} models, which
are often used in empirical analysis. 
However, they fail to capture the property of long memory in volatility.
To incorporate long memory in volatility, studies have proposed 
several models, such as long memory stochastic volatility\citep{breidt1998detection,harvey2007long}, 
fractionally integrated ARCH\citep{baillie1996fractionally}, 
and fractional stochastic volatility (FSV)\cite{comte1998long}.
The FSV model uses fractional Brownian motion with the Hurst parameter greater than $1/2$, which ensures long memory. 

Recently, \citet{gatheral2018volatility} analyzed log-volatility using the realized volatility(RV) 
as a proxy of true volatility 
and claimed that the time series of the log-volatility increments for stock and bond prices show  
rough behavior, that is, the Hurst exponent is smaller than $1/2$. 
They also claim that the time series shows monofractal behavior. 
From these empirical observations and the requirement of a small Hurst exponent for at-the-money skew\citet{fukasawa2011asymptotic},
they considered the log-volatility model by a fractional Brownian motion
with $H<1/2$, which is a variant of the FSV model and referred to as rough fractional stochastic volatility(RFSV) models.
Further empirical studies confirmed the roughness of the log-volatility
for thousands of stocks\citep{bennedsen2016decoupling} and implied volatility\citep{livieri2018rough}.

This study aims to provide further evidence of roughness of log-volatility in Bitcoin.
Many studies have investigated the statistical properties of Bitcoin, showing 
that stylized facts are also present in Bitcoin returns, see for example, \citet{bariviera2017some,chu2015statistical,takaishi2018}.
In this study, we use the multifractal detrended fluctuation analysis (MF-DFA) \citep{kantelhardt2002multifractal} 
to calculate the generalized Hurst exponent $h(q)$ of the log-volatility increments.
In the MF-DFA, $h(q)$ is obtained from the exponent of $q$th order fluctuation function. 
\cite{gatheral2018volatility} calculate $h(q)$ from the $q$th order structure function(SF) in a range of $q=(0,3]$
and find that $h(q)$ is constant, indicating that the time series is monofractal.
Here, we calculate $h(q)$ in a wide range of $q=[-25,25]$ using the MF-DFA and
investigate whether $h(q)$ is independent of $q$. 
In fact, we find evidence that $h(q)$ varies with $q$, which shows the multifractal nature of the log-volatility increments.

Section 2 in this letter describes the data and methodology, while
Section 3 presents the empirical results and Section 4 concludes.

\section{Data and Methodology}

In this study, we use Bitcoin Tick data (in dollars) traded on COINBASE
from January 28, 2015 to January 6, 2019 
and downloaded from Bitcoincharts\footnote{http://api.bitcoincharts.com/v1/csv/}. 
These data are used to construct the RV\citep{andersen1998answering,barndorff2001non,mcaleer2008realized} and
we use the RV as a proxy of volatility.
Let $p_{t_n}; t_n= n\delta t;  n=0,1,..,M$ be the $n$th Bitcoin prices
with sampling period $\delta t$ on day $t$, where $M=1440$min/$\delta t$min.
We define the return $r_{t,t_j}$ by the logarithmic price difference, namely,
\be
r_{t,t_j}=\log p_{t,t_j} -\log p_{t,t_{j-1}}.
\ee
The daily RV on day $t$ with sampling period $\delta t$ is given by
\be
RV^{\delta t}_t= \sum_{j=1}^M r_{t,t_j}^2.
\ee
In an ideal situation, in the limit of $\delta t \rightarrow 0$, the RV is expected to converge to the integrated volatility 
\be
IV_t = \int_t^{t+1440min} \sigma^2(\mu) d\mu,
\ee
where $\sigma(\mu)$ is the spot volatility.
Usually, the ideal situation is violated by the market microstructure noise (MMS), which
has many sources, such as the discreteness of the price, 
the bid-ask bounce, and properties of the trading mechanism.
The existence of MMS biases the RV, and 
especially, the bias strongly dominates at high frequency,
which can be visualized by the volatility signature plot\citep{andersen2000great}.
Here, we use a moderate 5-min sampling frequency to avoid strong bias at high-frequency 
by maintaining reasonable accuracy\citep{bandi2006separating,liu2015does}.
 
To obtain a more accurate RV, we could introduce a modification factor, such as the Hansen-Lunde (HL) factor\citep{hansen2005forecast}.
The HL factor is a multiplicative factor that corrects the RV, so that the average of RV matches the daily return variance.
Since the multiplicative factor does not change the Hurst exponent of the log volatility increments here, 
we use unmodified RV in our analysis.

To estimate the generalized Hurst exponent, we use the MF-DFA, 
which may be applied to non-stationary time series \citep{kantelhardt2002multifractal}. 
The MF-DFA has become a popular method to study the multifractal properties of various time series, 
and studies on Bitcoin have already applied this method, for example, \citet{takaishi2018,el2018bitcoin}.
Let us consider the time series $x_i:i=1,2,...N$.  
The MF--DFA consists of the following steps. 

(i) Determine the profile $Y(i)$,
\be
Y(i)=\sum_{j=1}^i (x_j- \bra x \ket),
\ee
where $\bra x \ket$ stands for the average of $x_i$.

(ii) Divide the profile $Y(i)$ into $N_s$ non-overlapping segments of equal length $s$, where $N_s \equiv {int} (N/s)$.
Since the length of the time series is not always a multiple of $s$, a short time period at the end of the profile may remain.
To utilize this part, we repeat the same procedure starting from the end of the profile.
Therefore, in total, we obtain $2N_s$ segments.

(iii) Calculate the variance
\be
F^2(\nu,s)=\frac1s\sum_{i=1}^s (Y[(\nu-1)s+i] -P_\nu (i))^2,
\ee
for each segment $\nu, \nu=1,...,N_s$ and
\be
F^2(\nu,s)=\frac1s\sum_{i=1}^s (Y[N-(\nu-N_s)s+i] -P_\nu (i))^2,
\ee
for each segment $\nu, \nu=N_s+1,...,2N_s$.
Here, $P_\nu (i)$ is the fitting polynomial to remove the local trend in segment $\nu$;
we use a cubic order polynomial.

(iv) Average over all segments and obtain the $q$th order fluctuation function
\be
F_q(s)=\left\{\frac1{2N_s} \sum_{\nu=1}^{2N_s} (F^2(\nu,s))^{q/2}\right\}^{1/q}.
\label{eq:FL}
\ee

(v) Determine the scaling behavior of the fluctuation function.
If the time series $r_{t_i}$ are long-range power law correlated,
$F_q(s)$ is expected to be the following functional form for large $s$.
\be
F_q(s) \sim s^{h(q)}.
\label{eq:asympto}
\ee
The scaling exponent $h(q)$ is called the generalized Hurst exponent.
$h(2)$ corresponds to the usual Hurst exponent.

We also determine $h(q)$ by the SF method used in \citet{gatheral2018volatility}.
The SF or moments $m(q,\Delta)$ of the log-volatility increments is defined as 
\be
m(q,\Delta)=\frac1N^{\prime} \sum_{k=1}^{N^{\prime}} |log(\sigma_{k\Delta})-log(\sigma_{(k-1)\Delta})|^q ,
\ee
and we expect the following relationship,
\be
m(q,\Delta) \sim c_q \Delta^{\zeta(q)},
\ee
where the exponent $\zeta(q)$ is assumed as $\zeta(q)=qh(q)$.
$h(q)$ can be obtained through $\zeta(q)$.

\section{Empirical Results}

In Figure 1(left), we show the time series of the RV calculated at $\delta t=$ 5min. 
Using the RV, the log-volatility on day $t$ is defined by $log(\sigma_t) \equiv log(RV_t^{1/2})$.
Then the log-volatility increments $LV_t^{\Delta}$ with $\Delta$ separation 
is defined by $LV_t^{\Delta}=log(\sigma_{t \Delta })-log(\sigma_{(t-1)\Delta})$.
Figure 1(right) shows the time series of $LV_t^{\Delta}$ at $\Delta=$1 day.

In the MF-DFA, $h(q)$ is obtained 
as the exponent of the fluctuation function $F_q(S)$.
Figure 2 shows $F_q(S)$ as a function of $s$ in the log-log plots.
We obtain $h(q)$ by fitting $F_q(S)$ to a linear function in $s=[80,280]$.
Figure 3 shows the results of $h(q)$ 
and clearly, $h(q)$ is smaller than 1/2, which indicates that the time series is anti-persistent, that is, rough.
The Hurst exponent $h(2)$ is estimated to be 0.144. This value is similar to those obtained for other assets\citep{gatheral2018volatility,bennedsen2016decoupling,livieri2018rough}.
Table 1 lists several selected values of $h(q)$.

\begin{figure}
%\vspace{5mm}
\centering
\includegraphics[height=4.1cm]{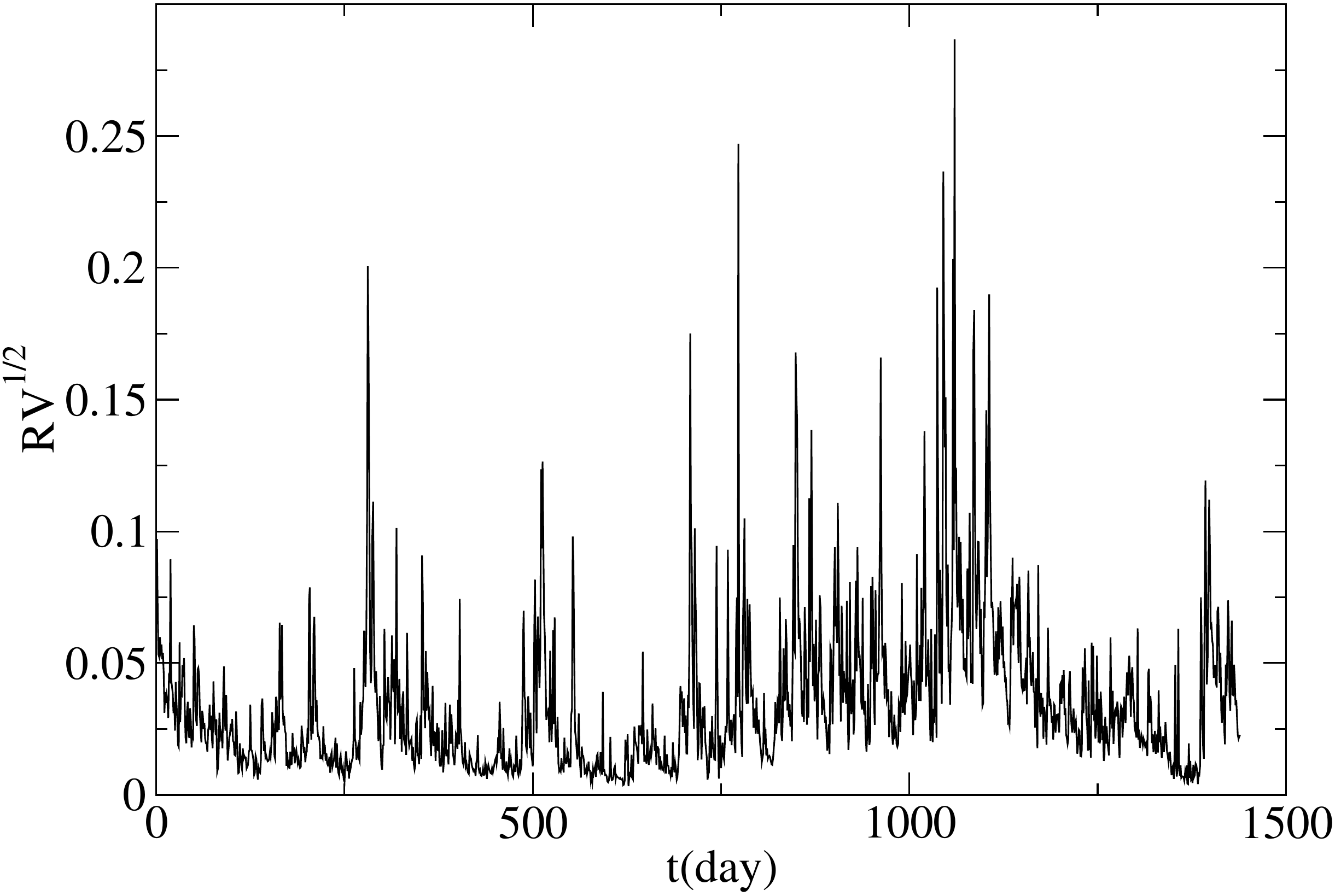}
\includegraphics[height=4.1cm]{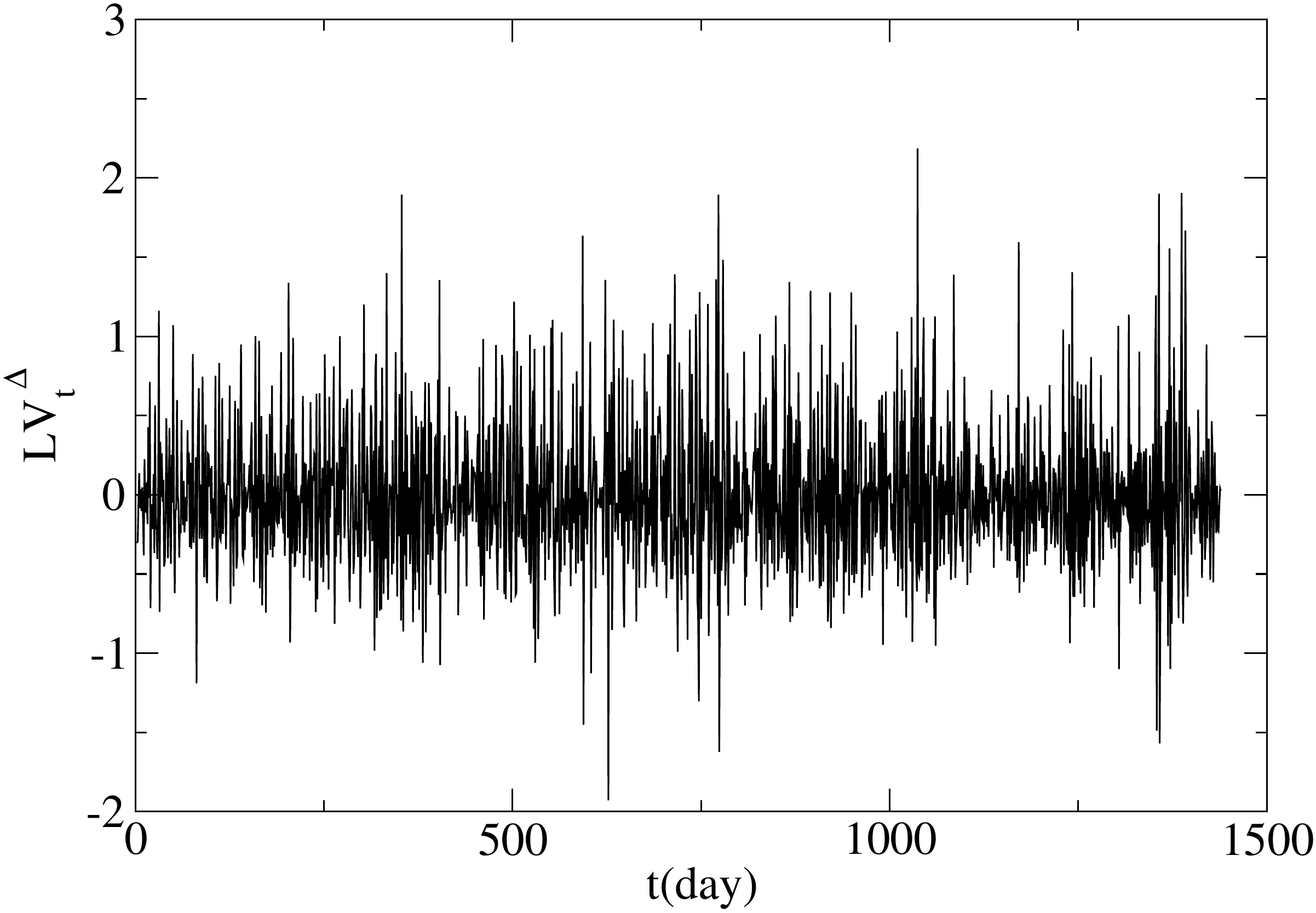}
\caption{(left): Time series of the RV at $\delta=5$min.
(right): Time series of the log-volatility increments $LV_t^\Delta$ at $\Delta=$1 day.
}
%\vspace{-2mm}
\end{figure}

\begin{figure}
%\vspace{5mm}
\centering
\includegraphics[height=6.cm]{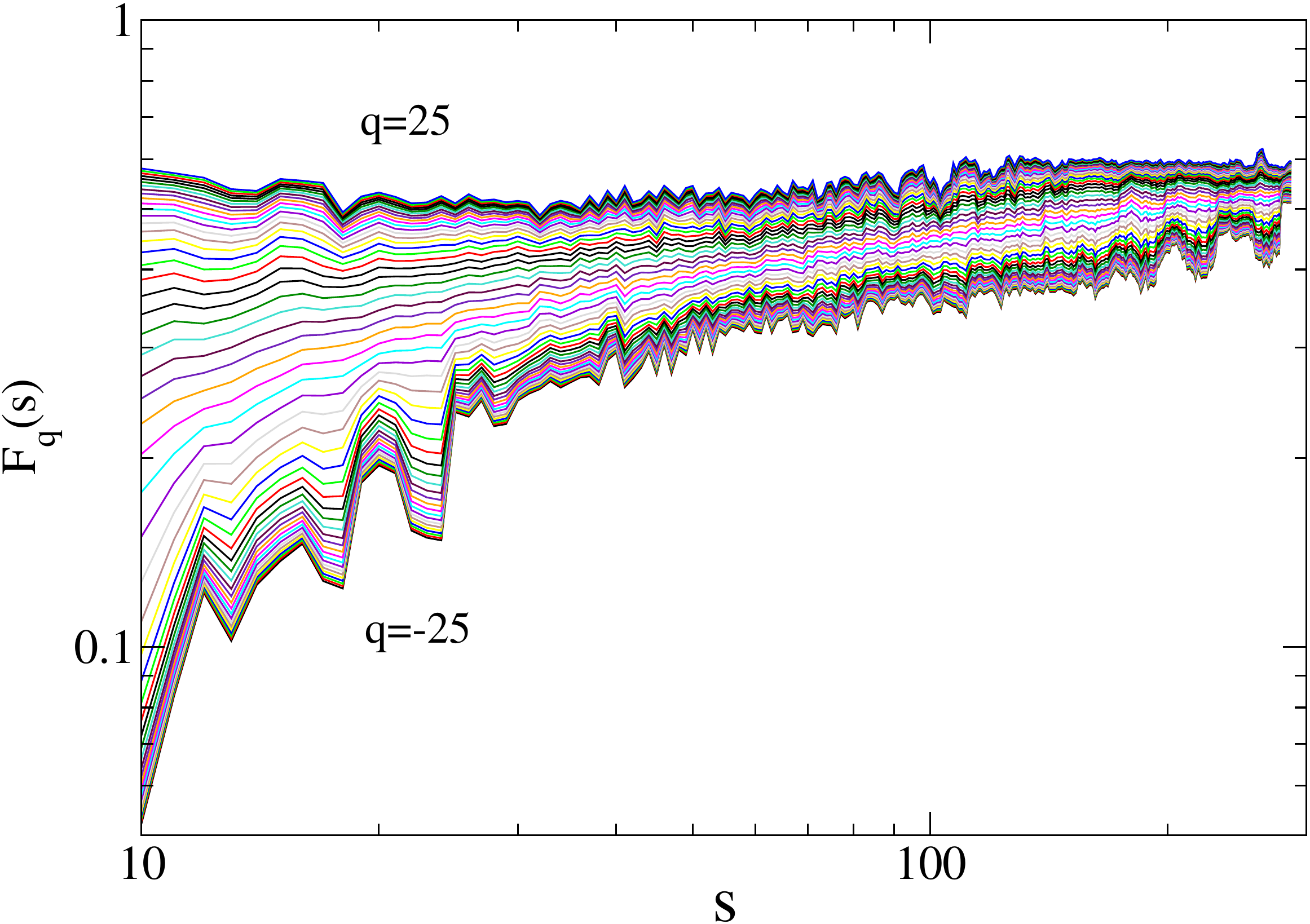}
\caption{Fluctuation function $F_q(s)$.
}
%\vspace{-2mm}
\end{figure}

\begin{figure}
%\vspace{5mm}
\centering
\includegraphics[height=6.cm]{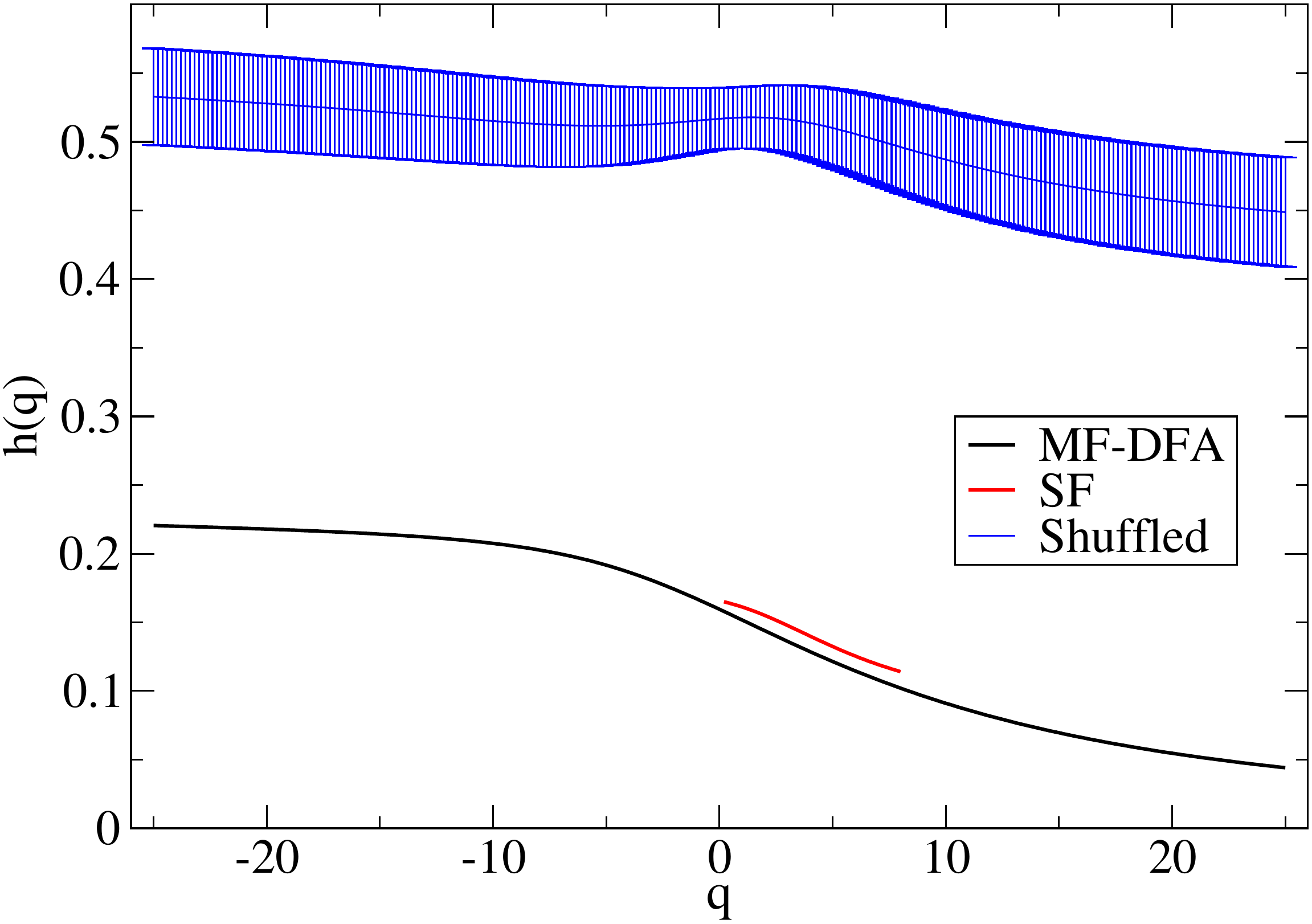}
\caption{
Generalized Hurst exponent $h(q)$ for the log-volatility increments of Bitcoin.
}
%\vspace{-2mm}
\end{figure}

\begin{table}
%\vspace{-5mm}
%\hspace{-10mm}
\centering
\caption{
Results of $h(q)$ from the MF-DFA and the SF method. The results indicated by ``Shuffled'' show $h(q)$ from 
20 shuffled time series.}
\hspace{-10mm}
\footnotesize
\begin{tabular}{c|ccccccc}
\hline
Bitcoin               &  h(-1) & h(0.2)  & h(1) & h(1.6) & h(2) & h(3) &  h(4)    \\
\hline
MF-DFA         & 0.167 & 0.158 & 0.152 &  0.147 & 0.144 & 0.136 & 0.129 \\
SF             & ---       & 0.165 & 0.161 &  0.159 & 0.155 & 0.148 & 0.140  \\
Shuffled       & 0.515(21) & 0.517(20) & 0.517(20) & 0.517(20) & 0.516(21) & 0.515(21)  & 0.512(22) \\
\hline
SPX            &  h(-1) & h(0.2)  & h(1) & h(1.6) & h(2) & h(3) &  h(4)    \\
\hline
MF-DFA         & 0.142 & 0.139  &  0.136 & 0.134 & 0.133 & 0.129 & 0.125 \\
SF             & ---   & 0.143  &  0.141 & 0.139 & 0.137 & 0.133 & 0.129 \\
Shuffled       & 0.486(18) & 0.485(19) & 0.485(19) & 0.484 (19) & 0.484(19) & 0.483(19) & 0.482(19) \\
\hline
\end{tabular}
%\vspace{-8mm}
\end{table}

In addition to roughness, we recognize that $h(q)$ varies as a function of $q$ ,
which is evidence of multifractality in the time series of 
the log-volatility increments.
We also calculate $h(q)$ using the SF method and
Figure 4 displays $m(q,\Delta)$.
We restrict the parameter $q$ in a range of $q=(0,8]$, 
since the SF becomes extremely noisy for $q>8$ within the current statistics. 
We obtain $h(q)$ by making a linear fit in the range of $\Delta=[1,40]$.
The results of $h(q)$ are plotted in Figure 3 together with those from the MF-DFA and
we find that the results are consistent with those from the MF-DFA.  

\begin{figure}
%\vspace{5mm}
\centering
\includegraphics[height=6cm]{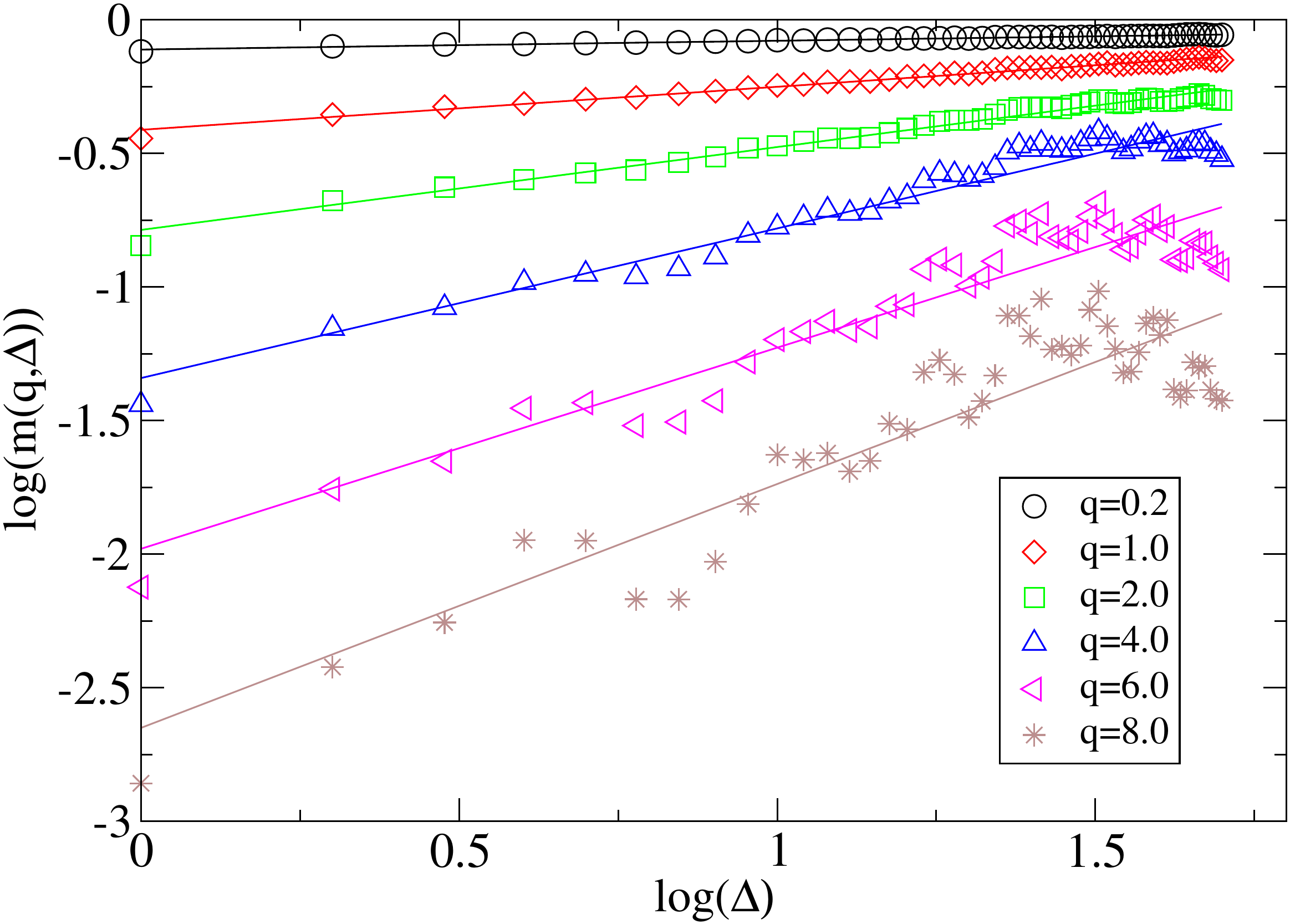}
\caption{
The structure function $m(q,\Delta)$, as a function of $\Delta$.
For better visibility, we have shifted some results vertically.
}
%\vspace{-2mm}
\end{figure}

The sources of multifractality are examined in \citet{kantelhardt2002multifractal}, who claim that
two sources contribute to the appearance of multifractality:
(i)temporal correlations and (ii)broad distributions. 
The distributions of the log-volatility increments are 
found to be close to Gaussian\citep{gatheral2018volatility,livieri2018rough}.
However, \citet{bennedsen2016decoupling} claim that non-Gaussian behavior of log-volatility is observed 
for a significant number of stocks. 
Within limited statistics, it is difficult to confirm Gaussian for 
our data set. Figure 6(left) depicts the distribution of log-volatility increments of Bitcoin.
Although it seems close to Gaussian, the kurtosis is calculated to be 4.4, which is greater than 3, the value of Gaussian.
Thus, it is possible that the distribution of log-volatility increments is slightly leptokurtic. 

\begin{figure}[ht]
%\vspace{5mm}
\centering
\includegraphics[height=4.cm]{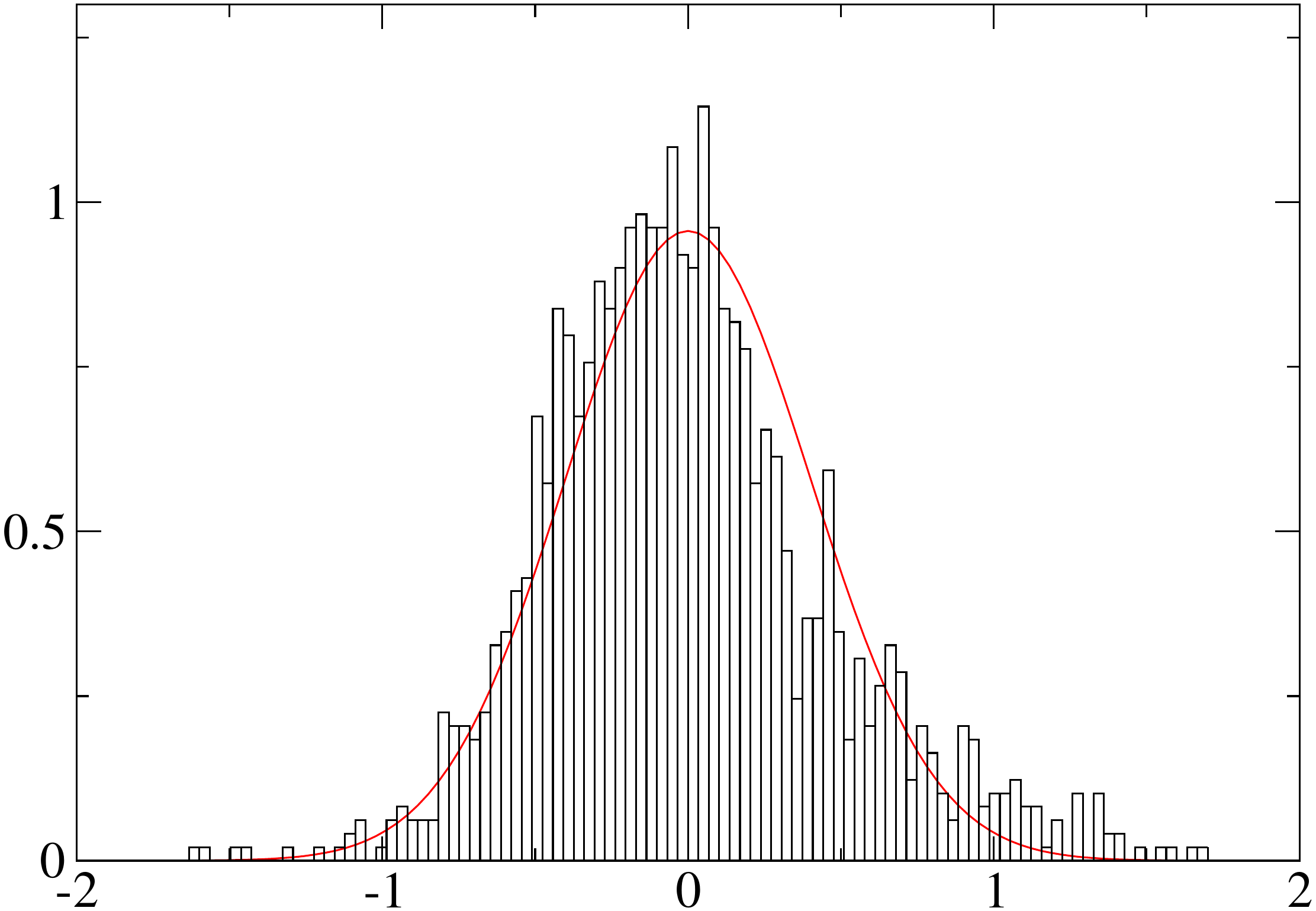}
\includegraphics[height=4.cm]{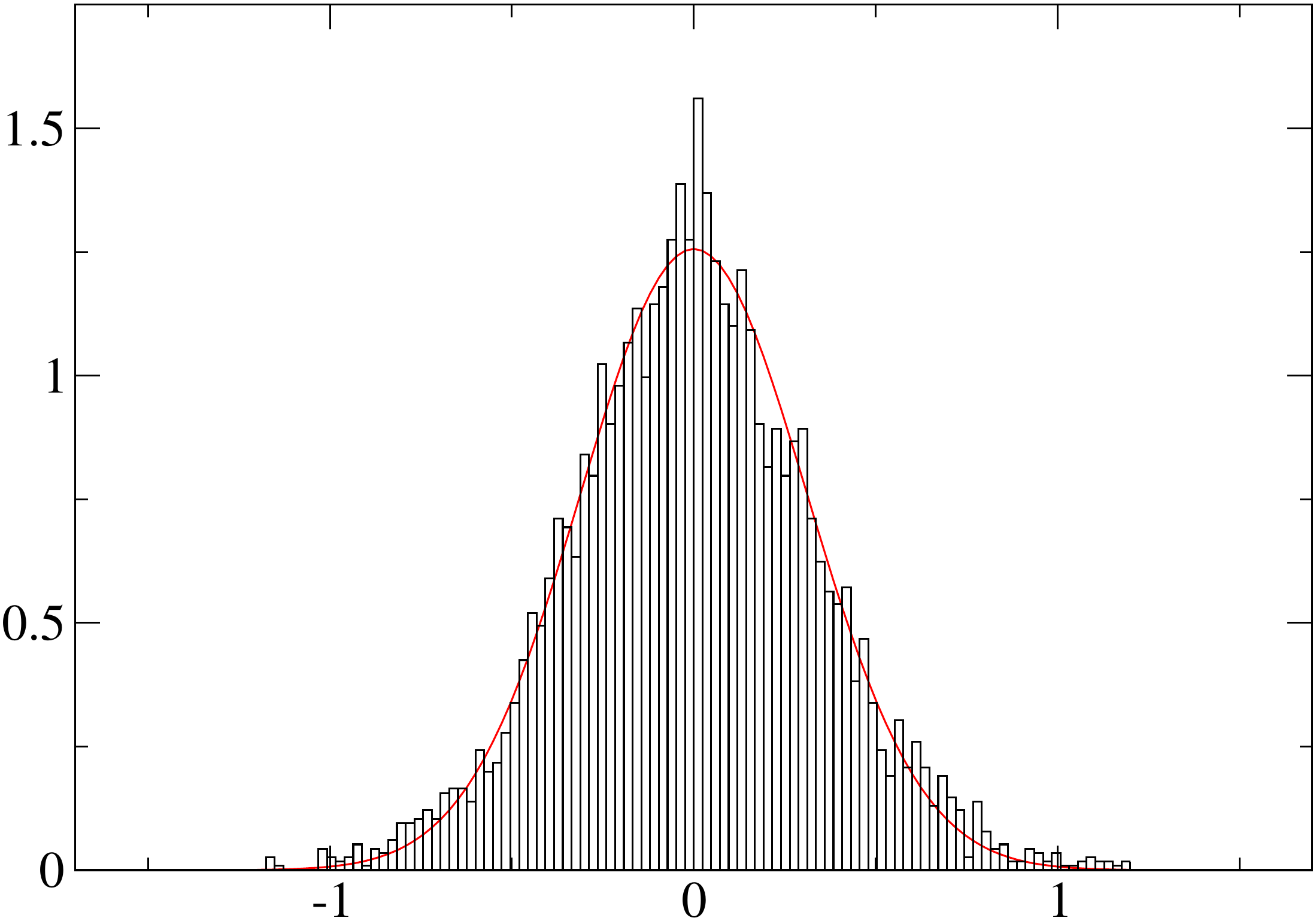}
\caption{
(left): Distribution of the log-volatility increments for Bitcoin.
(right): Distribution of the log-volatility increments for SPX.
The solid lines show a fit to Gaussian function.
The kurtosis is calculated to be 4.4(3.9) for Bitcoin(SPX)
}
%\vspace{-2mm}
\end{figure}

To investigate the origins of roughness and multifractality, 
we calculate $h(q)$ for the shuffled time series of the log-volatility increments.
The shuffling process can kill any temporal correlations; therefore, 
if both roughness and multifractality originate from temporal correlations,
we could expect that roughness and multifractality disappear for the shuffled time series.
In Figure 7, we show $h(q)$ from 20 shuffled time series of the log volatility increments, and
find that $h(q)$ comes close to 0.5. Since roughness disappears in the shuffling process,
it turns out that the temporal correlations contribute to roughness. 
On the other hand, it seems that $h(q)$ of the shuffled time series still varies slightly with $q$,
which means that the shuffled time series have weak multifractality.
To quantify the degree or strength of multifractality,
we measure $\Delta(h)=h(q_{min})-h(q_{max})$ used in \citet{zunino2008multifractal}. 
We also use the singularity spectrum $f(\alpha)$\citep{kantelhardt2002multifractal} defined by
\be 
\alpha = h(q)+qh^\prime(q),
\ee
\be
f(\alpha)=q[\alpha-h(q)] +1.
\ee
The range of variability of $\alpha$ in $f(\alpha)$, that is, $\Delta \alpha = max(\alpha)-min(\alpha)$,
also offers a degree of multifractality.
Figure 8 shows $f(\alpha)$  as a function of $\alpha$, and 
Table 2 lists the results of $\Delta h$ and $\Delta \alpha$.
Although both $\Delta h$ and $\Delta \alpha$ decrease for the shuffled time series,
they still remain finite. Thus, we conclude that the multifractality of log-volatility increments originates
partly from the distributional property of log-volatility increments, and
this observation supports the leptokurtic distribution for log-volatility increments. 

\begin{table}
\centering
\caption{
Results of $\Delta h$ and $\Delta \alpha$.
}
%\scriptsize
\hspace{-10mm}
\footnotesize
\begin{tabular}{c|cccc}
\hline
              & $\Delta h$ & $\Delta h$(Shuffled)  & $\Delta \alpha$  & $\Delta \alpha$(Shuffled)     \\
\hline
Bitcoin       & 0.232      &   0.099       &  0.232       &    0.155        \\
SPX           & 0.132      &   0.084       &  0.209       &   0.142      \\
\hline
\end{tabular}
%\vspace{-8mm}
\end{table}

Previously, the monofractal behavior of the log-volatility 
has been observed in a narrow range of $q$, that is, $q=(0,3]$. 
It might be difficult to identify the variability of $h(q)$ in such a narrow range.
We perform this same analysis for the Standard \& Poor's 500 Index (SPX) volatility
and try to obtain $h(q)$ in a wide range of $q$.
The 5min RV data of SPX from January 3, 2000 to February 27, 2019 are downloaded from the
Oxford-Man Institute of Quantitative Finance Realized Library\footnote{http://realized.oxford-man.ox.ac.uk/data/download}.
Figure 5 (right), 6, and 8 display the distribution of the log-volatility increments,
$h(q)$ and $f(\alpha)$, respectively.
Typically, the results are very similar to those of Bitcoin,
and importantly, we recognize the variability of $h(q)$, that is, multifractality for SPX.

\begin{figure}
%\vspace{5mm}
\centering
\includegraphics[height=6.cm]{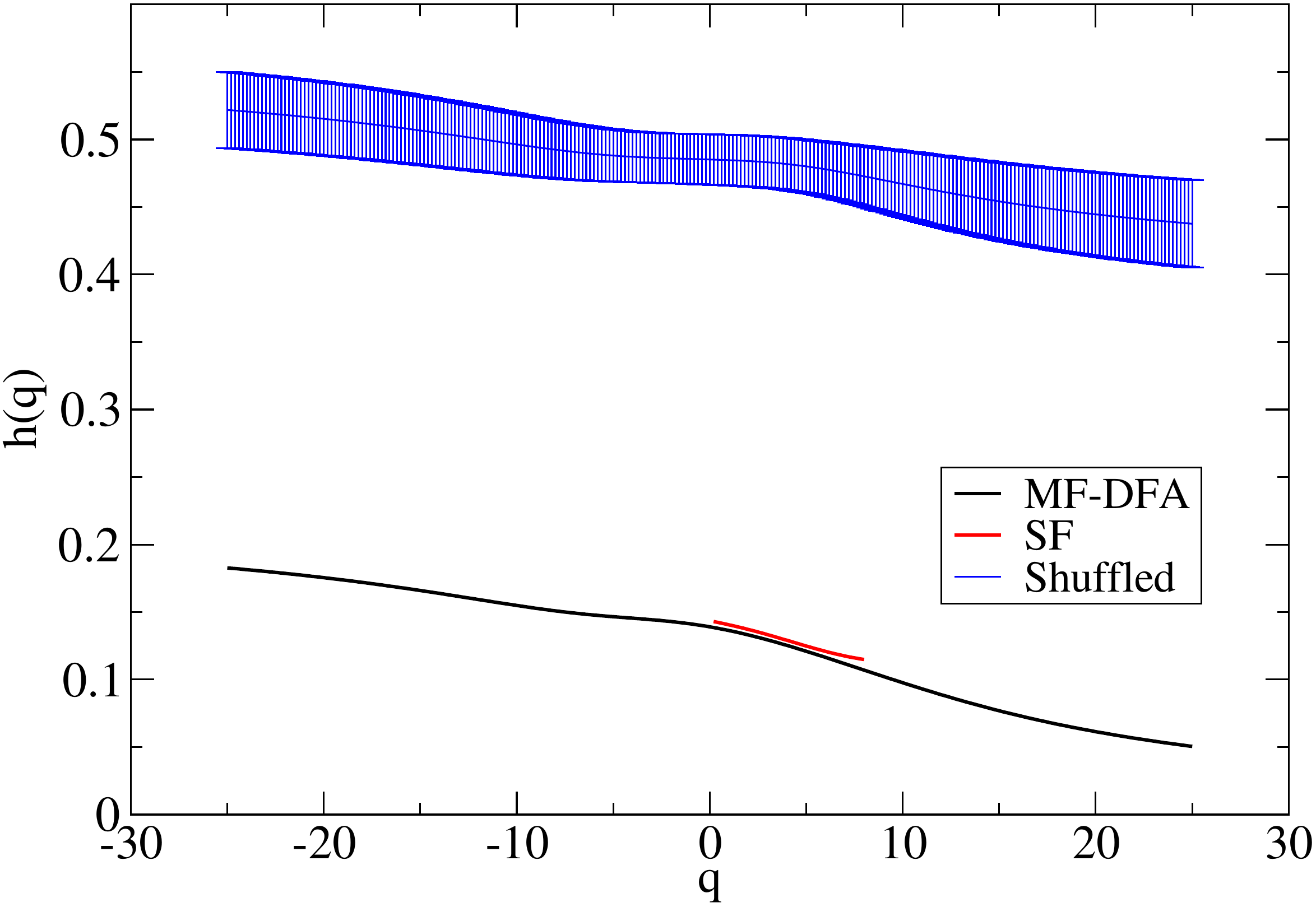}
\caption{
Generalized Hurst exponent $h(q)$ of the log-volatility increments for SPX.
}
%\vspace{-2mm}
\end{figure}

\begin{figure}
%\vspace{5mm}
\centering
\includegraphics[height=6.cm]{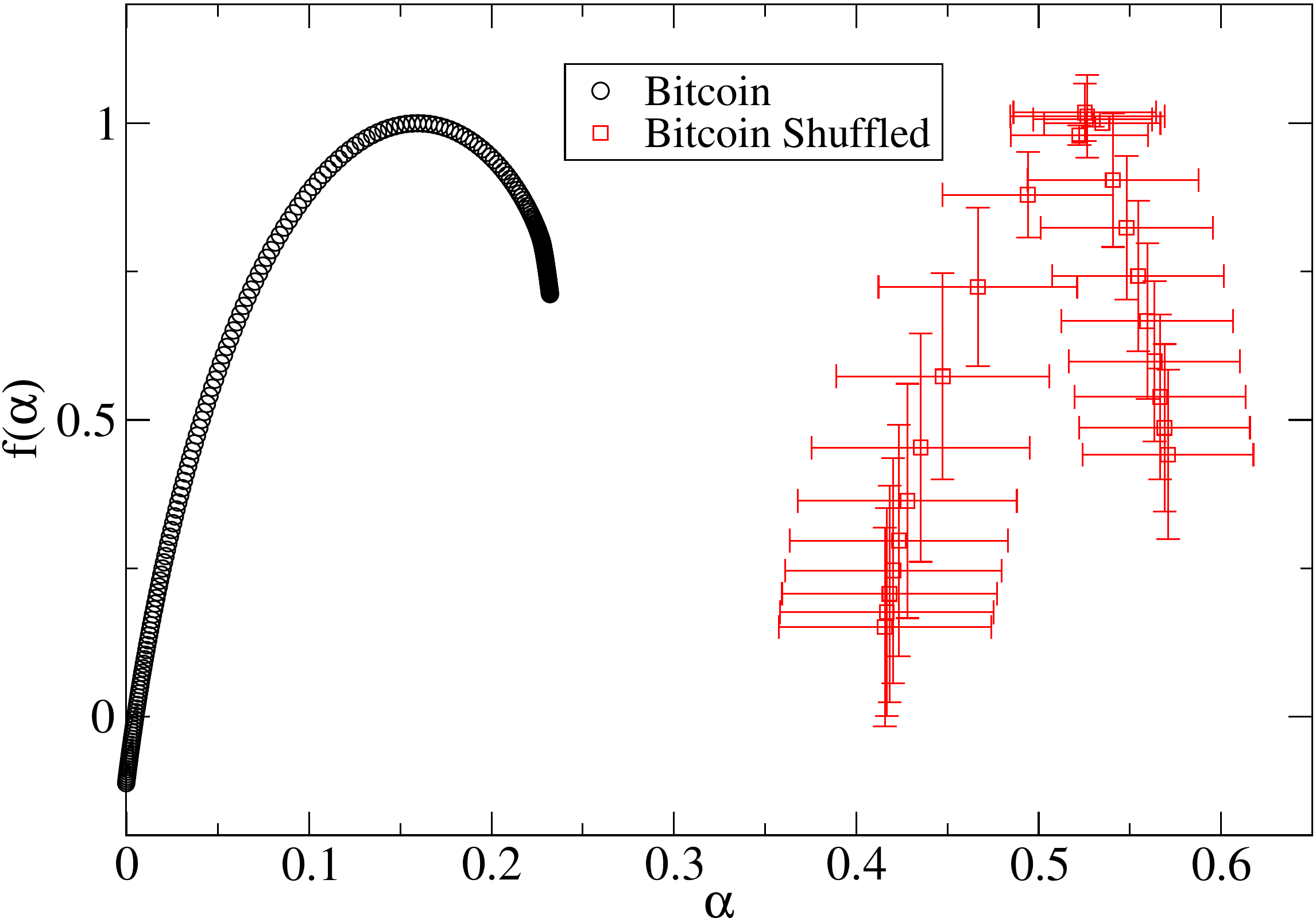}
\caption{
The singularity spectrum $f(\alpha)$ for Bitcoin. 
}
%\vspace{-2mm}
\end{figure}
\begin{figure}
%\vspace{5mm}
\centering
\includegraphics[height=6.cm]{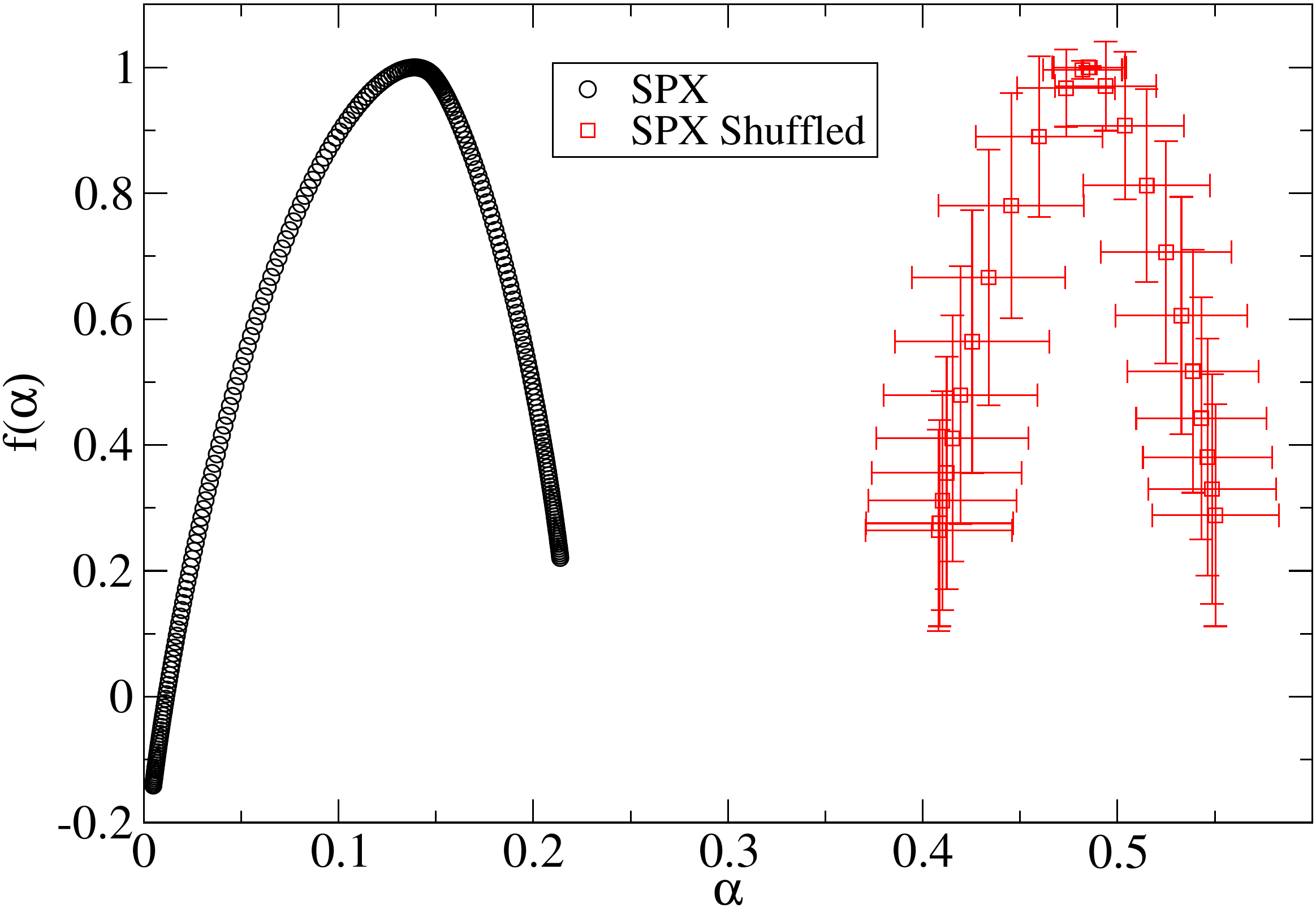}
\caption{
The singularity spectrum $f(\alpha)$ for SPX.
}
%\vspace{-2mm}
\end{figure}

\section{Conclusions}

We investigate the generalized Hurst exponent $h(q)$ of the log-volatility increments for Bitcoin
using the MF-DFA. 
We find that $h(q)$ is less than 1/2, which is consistent with the previous results empirically observed for 
other assets.
Furthermore, 
we also find that $h(q)$ varies with $q$, which indicates the existence of multifractality 
in the time series of the log-volatility increments. 
Using a shuffled time series, 
we confirm that while roughness is related to temporal correlations,
multifractality originates, in part, from the distributional properties of log-volatility increments.
From a rough volatility perspective, \citet{neuman2018fractional}
consider a fractional Brownian motion when the Hurst exponent goes to zero and
show that it converges to a Gaussian random distribution close to a log-correlated Gaussian field related 
to some multifractal processes\citep{mandelbrot1997multifractal}.
Our finding of the existence of multifractality in log-volatility increments 
supports a more serious consideration of such a volatility model, including multifractality.

\section*{Acknowledgment}
Numerical calculations for this work were carried out at the
Yukawa Institute Computer Facility and at the facilities of the Institute of Statistical Mathematics.
This work was supported by JSPS KAKENHI Grant Number JP18K01556 and
the ISM Cooperative Research Program（2018-ISMCRP
-0006).

\section*{References}

\bibliography{mybibfile}

\end{document}